\documentclass[10pt,aps,prl,twocolumn,amsmath,amssymb,superscriptaddress,showpacs,footinbib]{revtex4-1}

\usepackage{amsmath}
\usepackage{graphicx}
\usepackage{xcolor}

\begin{document}

\title{Tunable quantum interference in bilayer graphene in double-resonant Raman scattering}

\author{Felix Herziger}
\affiliation{Institut f\"ur Festk\"orperphysik, Technische Universit\"at Berlin, Hardenbergstrasse 36, 10623 Berlin, Germany}

\author{Christoph Tyborski}
\affiliation{Institut f\"ur Festk\"orperphysik, Technische Universit\"at Berlin, Hardenbergstrasse 36, 10623 Berlin, Germany}

\author{Oliver Ochedowski}
\affiliation{Fakult\"at f\"ur Physik und CENIDE, Universit\"at Duisburg-Essen, 47048 Duisburg, Germany}

\author{Marika Schleberger}
\affiliation{Fakult\"at f\"ur Physik und CENIDE, Universit\"at Duisburg-Essen, 47048 Duisburg, Germany}

\author{Janina Maultzsch}
\email{janina.maultzsch@fau.de}
\affiliation{Institut f\"ur Festk\"orperphysik, Technische Universit\"at Berlin, Hardenbergstrasse 36, 10623 Berlin, Germany}
\affiliation{Department f{\"u}r Physik, Friedrich-Alexander-Universit\"{a}t Erlangen-N\"{u}rnberg, Staudtstra\ss e 7, 91058 Erlangen, Germany}


\begin{abstract}
The line shape of the double-resonant $2D$ Raman mode in bilayer graphene is often considered to be characteristic for a certain laser excitation energy. Here, in a joint experimental and theoretical study, we analyze the dependence of the double-resonant Raman scattering processes in bilayer graphene on the electronic broadening parameter $\gamma$. We demonstrate that the ratio between symmetric and anti-symmetric scattering processes sensitively depends on the lifetime of the electronic states, explaining the experimentally observed variation of the complex $2D$-mode line shape. 
\end{abstract}

\maketitle%

\section{Introduction}
Establishing the concept of double-resonant Raman scattering by Thomsen and Reich in 2000\cite{PhysRevLett.85.5214} has led to a fundamental understanding of inelastic phonon scattering processes in graphene and related graphitic carbons \cite{Kurti2002,Saito2001,PhysRevB.70.155403,Ferrari2000}. It enabled a unified description of the complete Raman spectrum of these materials, explaining anomalous peak shifts with laser excitation energy and the occurence of various Raman modes in graphene, few-layer graphene, graphite, and carbon nanotubes \cite{PhysRevB.65.233402,PhysRevB.84.035433,PhysRevB.85.235447,PhysRevLett.97.187401,PhysRevB.87.075402,PhysRevB.78.125418,PhysRevB.90.245431,POPOV2015436}.   

The $2D$ mode is one of the most discussed peaks in the Raman spectrum of graphene and related graphitic carbons. In single-layer, graphene the $2D$ mode can empirically be fitted by a single Lorentzian line shape with a slight asymmetry towards the high-frequency side \cite{PhysRevB.80.165413,10.1021/nl400917e}. In bilayer graphene this mode splits up into four different components with different relative intensities and frequencies as a function of the laser excitation energy. The origin of these four components and their relation to the underlying double-resonant Raman scattering processes has been discussed by various authors in the past \cite{PhysRevB.79.125426,PhysRevLett.97.187401,PhysRevB.76.201401,10.1016/j.carbon.2010.11.053} and only recently a consistent interpretation of the complex $2D$-mode line shape has been presented \cite{PhysRevLett.113.187401}.

The consideration of interference effects is an integral part for the understanding of Raman spectra from graphene sheets. Only internal quantum interference between different Raman scattering processes explains the narrow line widths of the 2D band or the absence of phonons with \textbf{q}=\textbf{K} in the Raman spectrum of graphene~\cite{PhysRevB.70.155403}. The influence of the  line width broadening has been elucidated for single-layer graphene, influencing the intensity of the D mode and consequently the D/G intensity ratios as demonstrated by Bruna \textit{et al.}~\cite{10.1021/nn502676g} and Fr\"{o}hlicher \textit{et al.}~\cite{PhysRevB.91.205413}.

Here, we present a joint experimental and theoretical analysis of the $2D$-mode line shape in bilayer graphene as a function of the electronic broadening $\gamma$ in the double-resonance process. We demonstrate that the electronic broadening determines the interference of different scattering paths resulting in different intensity ratios of the Raman peaks related to symmetric and anti-symmetric scattering processes. We support our theoretical findings by experiments in which the electronic broadening is tuned as a function of defect densities and as a function of the electron-phonon coupling.

\section{Experimental and calculation details}
Freestanding bilayer graphene samples were prepared by micro-mechanical exfoliation from natural graphite crystals onto prepatterned SiO$_2$/Si substrates with an oxide thickness of 100\,nm. Holes in the substrate were fabricated by reactive ion etching (RIE), exhibiting a diameter of 3\,$\mu$m and depth of approximately 8\,$\mu$m. Defective bilayer graphene samples were prepared by swift heavy ion irradiation (Xe$^{26+}$, 91 MeV)\cite{Aumayr2011,Ochedowski2015} at normal incidence using different ion fluxes \cite{PhysRevB.90.245431}. 

Raman measurements were carried out with a Horiba HR\,800 spectrometer, equipped with solid-state and gas lasers. Raman spectra were recorded in back-scattering geometry under standard ambient conditions using a 1800\,lines/mm grating and an $100\,\times$ objective, yielding a spectral resolution of approximately 1\,cm$^{-1}$. The laser excitation energies were between 1.96\,eV and 2.81\,eV and the laser power was kept below 0.5\,mW during all measurements in order to avoid laser-induced modification of the graphene samples \cite{10.1038/srep02355, pssb.201552411}. 

The calculation of the double-resonant Raman scattering cross-section was done as proposed in Refs.~\cite{PhysRevLett.85.5214}, \cite{PhysRevB.84.035433}, and \cite{PhysRevLett.113.187401}. We used the $GW$-corrected electronic bandstructure and TO phonon dispersions for bilayer graphene from Ref.~\cite{PhysRevLett.113.187401}, which were both calculated using the Quantum Espresso DFT Code \cite{10.1088/0953-8984/21/39/395502,PhysRevLett.113.187401}. For simplicity, we assumed all matrix elements to be constant and restricted the integration to the one-dimensional $\Gamma-K-M$ high-symmetry direction. This approach is justified as the main contributions in the double-resonance process stem from the high-symmetry lines \cite{PhysRevB.84.035433, PhysRevB.87.075402, PhysRevB.90.245431}.

\begin{figure}
\includegraphics{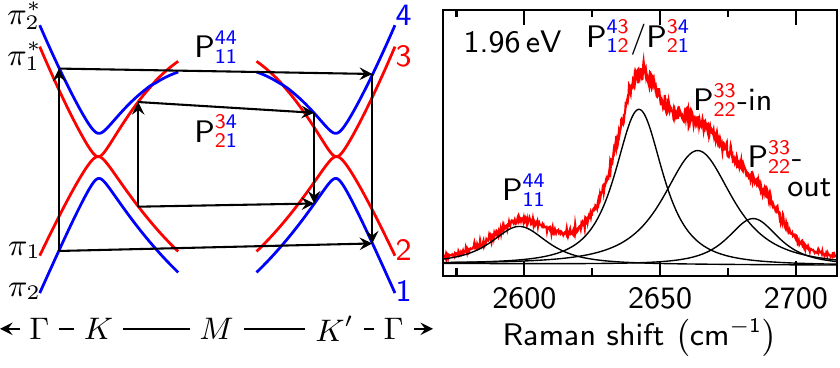}
\caption{Left: Illustration of the different intervalley double-resonant Raman scattering processes P$^{lj}_{mi}$ in the electronic bandstructure of bilayer graphene. The indices $m,i$ and $l,j$ number the valence and conduction bands, respectively. For symmetric scattering processes equals $m=i$ and $l=j$. For anti-symmetric processes is $m\neq i$ and $l\neq j$. Right: Representative Ramann spectrum of freestanding bilayer graphene at 1.96\,eV laser energy. The assignment between the scattering processes and the different spectral contributions is indicated. All contributions stem from inner processes except for the highest-frequency peak, indicated with P$^{33}_{22}$-out. \label{figDR}}
\end{figure}

\section{Results and discussion}
Each double-resonant intervalley scattering process P$^{lj}_{mi}$ in bilayer graphene can be described by four indices that refer to the band indices of the initial electron $m$ and the excited electron $l$, as well as the band indices of the scattered electron $j$ and the scattered hole $i$. Two representative examples of these scattering processes are shown in Fig.~\ref{figDR}. For symmetric processes, electrons and holes are scattered between bands with the same index, \textit{i.e.}, $m=i$ and $l=j$, whereas the band indices are changing during anti-symmetric scattering processes, \textit{i.e.}, $m\neq i$ and $l\neq j$. Furthermore, the scattering processes split up into so-called 'inner' and 'outer' processes according to the resonant phonon wave vectors in the double-resonance process \cite{PhysRevB.84.035433}. As demonstrated before \cite{PhysRevLett.113.187401}, the complex $2D$-mode line shape can be understood as composed of contributions from symmetric scattering processes P$^{44}_{11}$ and $P^{33}_{22}$ on the low- and high-frequency side, respectively, and contributions from degenerate anti-symmetric scattering processes (P$^{43}_{12}$, P$^{34}_{21}$) in the centre (compare Fig.~\ref{figDR}). Except for the highest-frequency peak (P$^{33}_{22}$-out in Fig.~\ref{figDR}), all contributions to the 2D mode stem predominantly from inner processes, \textit{i.e.} phonons with wave vector between $\Gamma$ and $K$\cite{PhysRevLett.113.187401,Mohr2010}
 
\begin{figure}
\includegraphics{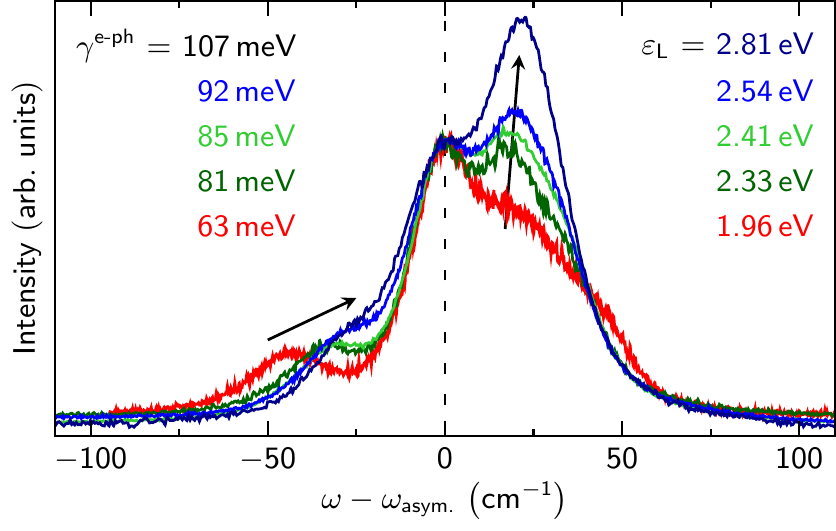}
\caption{Laser-energy dependence of the $2D$-mode line shape in freestanding bilayer graphene. All spectra are frequency- and intensity-normalized to the contribution from anti-symmetric scattering processes. $\gamma^{\text{e-ph}}$ is calculated according to Ref.~\cite{PhysRevB.84.035433}. \label{fig3} Raman spectra for $\varepsilon_{\text{L}}=$1.96\,eV, 2.33\,eV, 2.41\,eV, and 2.54\,eV are taken from Ref.~\cite{PhysRevLett.113.187401}}
\end{figure}

Although the assignment between the scattering processes and the different spectral features in the $2D$-mode line shape has been discussed before \cite{PhysRevLett.113.187401, PhysRevLett.97.187401, PhysRevB.76.201401, PhysRevB.77.245408, PhysRevB.80.241414, 10.1016/j.carbon.2010.11.053, 10.1021/nl300477n, 10.1002/jrs.2435}, the origin of their different relative intensities and their distinctive variations as a function of the laser excitation energy remained unclear. It is however an important aspect of the 2D-mode in bilayer graphene as can be seen in Fig.~\ref{fig3}. The relative intensities of single contributions drastically change as a function of the laser excitation energy.\newline
In this context, we want to show that the electronic broadening $\gamma$ plays a decisive role. The electronic broadening, which reflects the inverse lifetime of the electronic states, is usually approximated as given by Ref.~\cite{PhysRevB.84.035433} and is composed of three different contributions
\begin{equation}
\gamma = \gamma^{\text{e-ph}} + \gamma^{\text{D}} + \gamma^{\text{e-e}}. \label{eq:Gamma}
\end{equation}
Here, $\gamma^{\text{e-ph}}$ reflects the intrinsic contribution from electron-phonon scattering, which can be further separated into the contribution from $\Gamma$ and $K$ point phonons \cite{PhysRevB.80.165413,PhysRevB.84.035433}. $\gamma^{\text{D}}$ refers to the electron-defect scattering rate and $\gamma^{\text{e-e}}$ represents the contribution to $\gamma$ from scattering between charge carriers. In principle, all three contributions can be manipulated experimentally. For instance, $\gamma^{\text{e-ph}}$ depends on the laser excitation energy $\varepsilon_L$ \cite{PhysRevB.84.035433}, as well as on temperature \cite{10.1021/nl402696q, PhysRevB.90.125414}. Naturally, $\gamma^{\text{D}}$ depends on the amount and type of defects and is also influenced by the laser energy \cite{PhysRevB.84.035433,PhysRevB.91.205413}. Finally, $\gamma^{\text{e-e}}$ can be tuned by shifting the Fermi level E$_{\text{F}}$. In fact, Basko \textit{et al.}\cite{PhysRevB.80.165413} demonstrated that $\gamma^{\text{e-e}}$ scales linearly with $\vert E_{\text{F}}\vert$ in single-layer graphene. Previous works on single-layer graphene already extensively investigated the effect of defects and doping on $\gamma$ in the Raman scattering process by analyzing the $D/G$ and $2D/G$ mode intensity ratios \cite{10.1016/j.carbon.2009.12.057, 10.1021/nl201432g, PhysRevB.88.035426, 10.1021/nl300901a, 10.1021/nn502676g, PhysRevB.91.205413}. For bilayer graphene, different studies that analyze the doping dependence of the $E_{\text{2g}}$ and $E_{\text{u}}$ $\Gamma$ point vibrations were reported \cite{PhysRevLett.101.136804, PhysRevB.79.155417, PhysRevB.80.155422, PhysRevB.80.241417}. However, the influence of $\gamma$ on the line shape of double-resonant Raman modes in bilayer graphene has not been considered before. In the following we will discuss the influence of the broadening factor $\gamma$ theoretically and will report corresponding experimental Raman spectra. 
\begin{figure}
\includegraphics{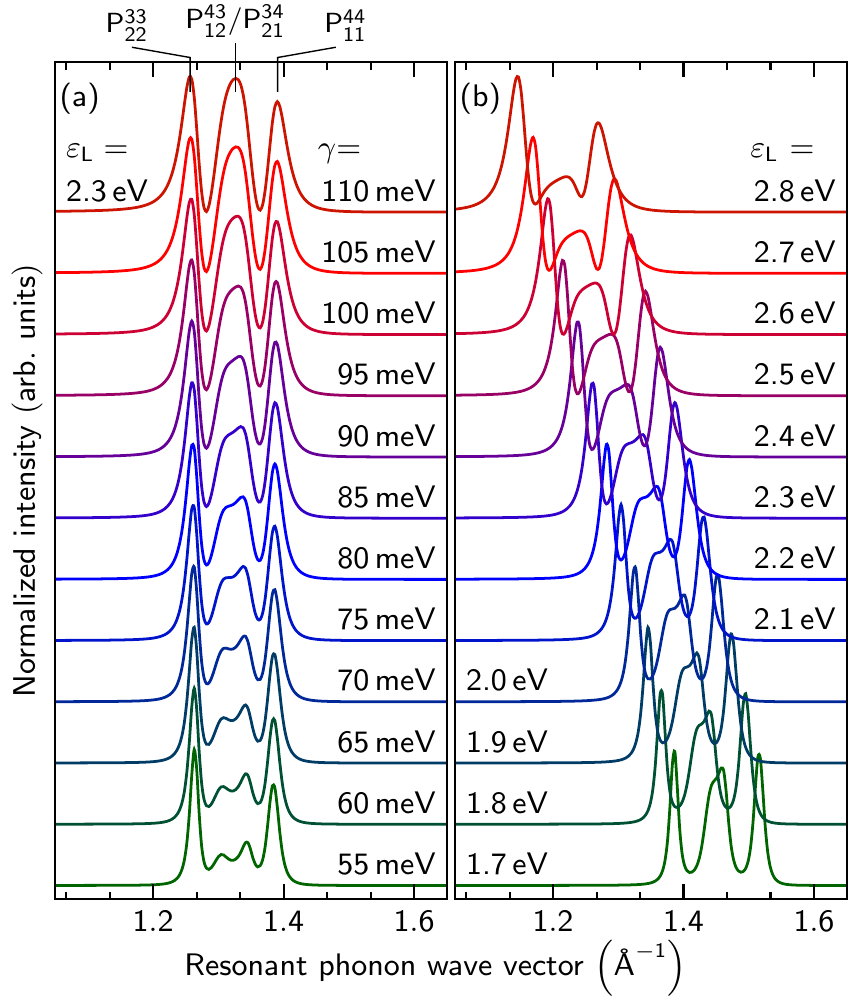}
\caption{Calculated resonant phonon wave vectors for inner processes along the K - M direction in the double-resonant $2D$-mode scattering process of bilayer graphene as a function of (a) the electronic broadening $\gamma$ and (b) the laser excitation energy $\varepsilon_L$ ($\gamma$ varies as described in Ref.~\cite{PhysRevLett.113.187401}). All spectra are normalized to the contribution from the P$^{33}_{22}$ scattering process and are vertically offset for clarity. \label{fig1}}
\end{figure}

\begin{figure*}
\includegraphics{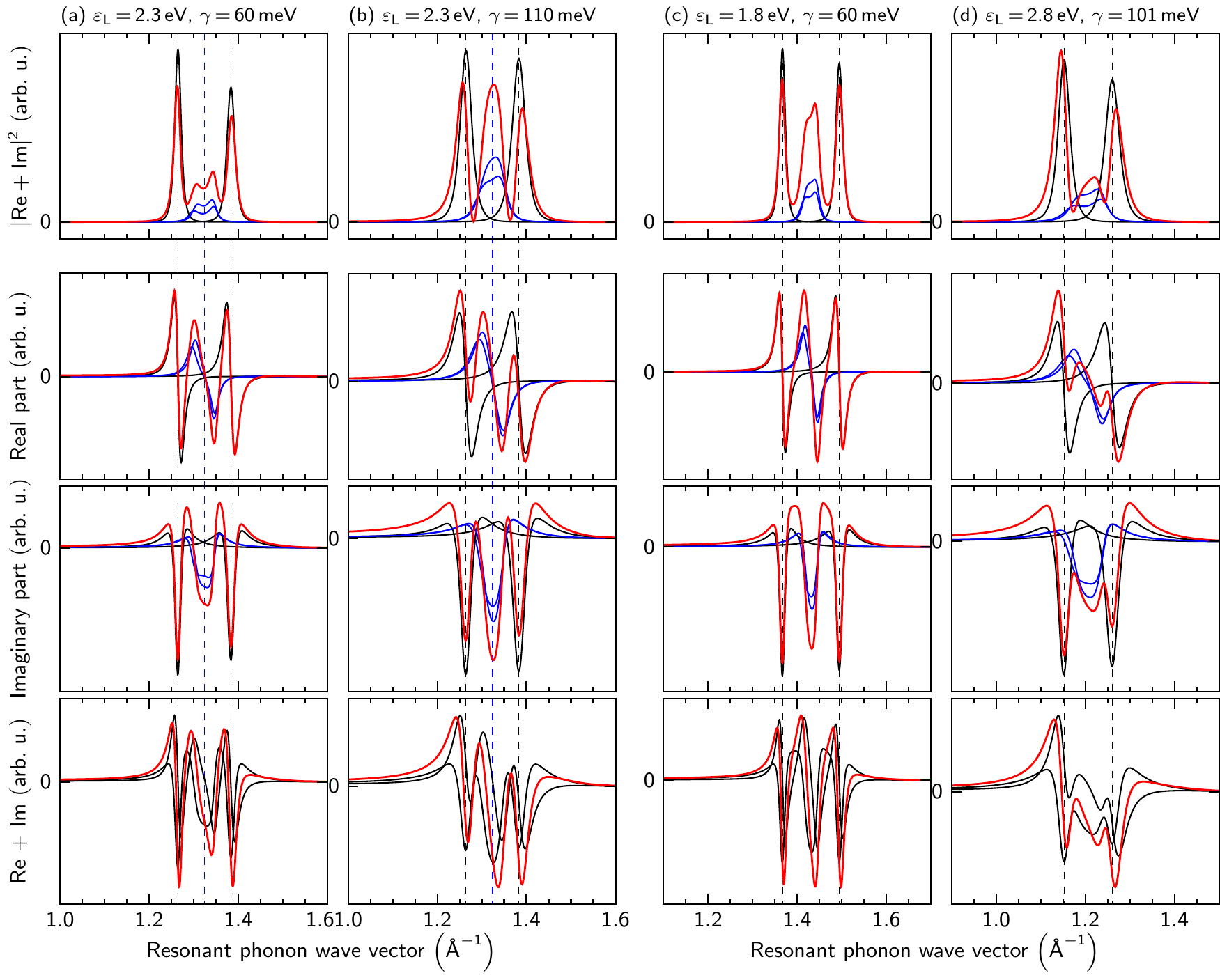}
\caption{Decomposition of the resonant phonon wave vectors along $\Gamma-K-M$ (top row) into the corresponding real and imaginary parts (middle rows) for different calculational scenarios: (a),(b) variation of the electronic broadening $\gamma$ at constant laser energy $\varepsilon_L$; (c),(d) simultaneous variation of both $\gamma$ and $\varepsilon_L$. Rows 1 to 3: Black lines refer to the symmetric scattering processes, blue lines refer to the asymmetric processes, and red lines represent the sum of the four different contributions. Row 4: Black lines refer to the sum of the real part and the imaginary parts of the scattering amplitude, respectively. Red lines refer to the sum of all respective contributions. Dashed vertical lines indicate the resonant phonon wave vectors for symmetric and anti-symmetric scattering processes P$^{33}_{22}$ and P$^{44}_{11}$ (compare Fig.~\ref{fig1}). \label{fig2}}
\end{figure*}

Figure \ref{fig1} presents calculations of the one-dimensional resonant phonon wave vectors for the so-called inner processes as a function of the electron broadening (a) and the laser excitation energy (b). For reasons of simplicity, we only discuss "inner" processes, but all statements are generally valid for "outer" processes. The correspondence between the scattering processes P$^{lj}_{mi}$ and different resonant phonon wave vectors is given at the top of this plot. We distinguish between two situations: Fig.~\ref{fig1}\,(a) presents the evolution of the resonant phonon wave vectors at a constant laser excitation energy $\varepsilon_L$ with different electronic broadenings. Fig.~\ref{fig1}\,(b) shows the variation of the phonon wave vectors by changing $\varepsilon_L$ and $\gamma$ simultaneously according to the equation given in Ref.~\cite{PhysRevLett.113.187401}.
Figs.~\ref{fig1} (a) and (b) show different dependencies of the intensity ratios between symmetric (P$^{33}_{22}/\text{P}^{44}_{11}$) and anti-symmetric (P$^{43}_{12}/\text{P}^{34}_{21}$) scattering processes. As can be seen from Fig.~\ref{fig1}\,(a), an increasing $\gamma$ at constant $\varepsilon_L$ leads to an enhancement of the relative intensities of anti-symmetric processes as compared to the contribution from symmetric processes. In contrast, by simultaneously increasing $\varepsilon_{\text{L}}$ and $\gamma$, we observe the opposite effect [see Fig.~\ref{fig1}\,(b)], \textit{i.e.}, the relative intensities from anti-symmetric processes decrease although the broadening $\gamma$ increases. Further, we would like point out the significant line shape variations of the symmetric contributions in both calculations for an increasing value of $\gamma$. We observe a strong asymmetry of  P$^{44}_{11}$ and P$^{33}_{22}$ towards the high- and low- wave vector side, respectively. Such asymmetric line shapes are, for instance, observed in Fano resonances and are usually identified with interference effects. \newline
\begin{figure}
\centering
\includegraphics{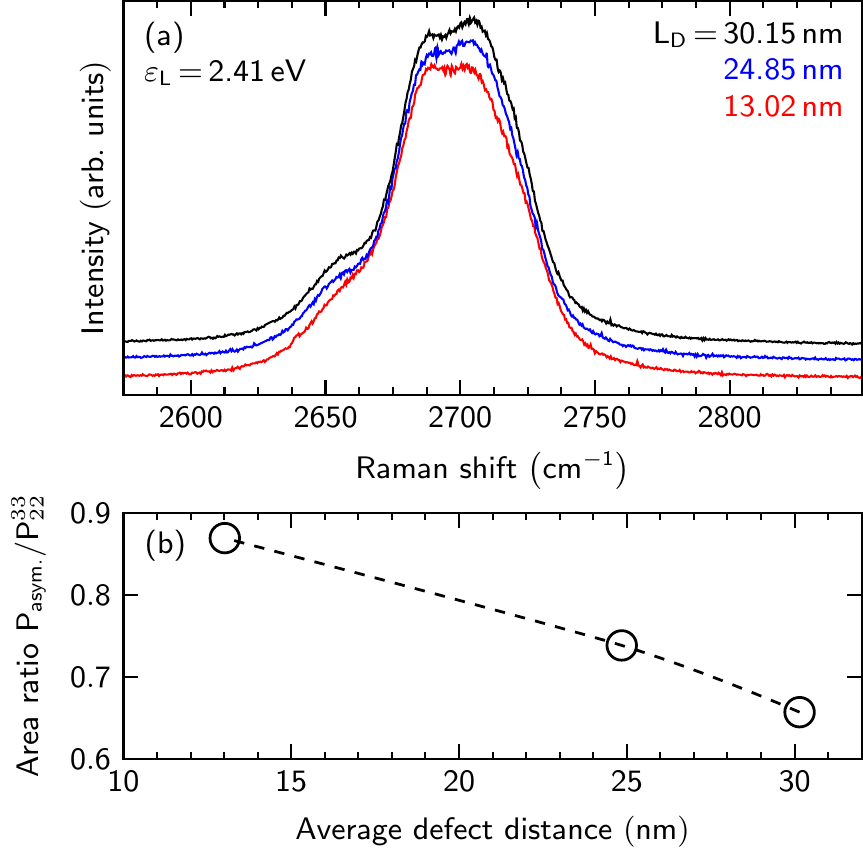}%
\caption{(a) Raman spectra of the $2D$ mode of ion-irradiated bilayer graphene. The average defect distances $L_D$ for each sample are given next to the spectra. (b) Evolution of the intensity ratio between the symmetric and anti-symmetric contributions as a function of the average defect distance.
\label{fig:BilayerDefects}}%
\end{figure}  
In order to understand the calculated spectra in Fig.~\ref{fig1} more thoroughly, we decompose the calculated resonant phonon wave vectors into their real and imaginary parts. This decomposition is shown in Figs.~\ref{fig2}\,(a) and (b) for a calculation with constant laser excitation energies and a variation of the electronic broadening as well as for a calculation with different laser excitation energies and different electronic broadening factors in Fig.~\ref{fig2}\,(c) and (d). The contributions from the single scattering processes P$^{lj}_{mi}$ are shown in black, their sum is represented by red lines. First, we want to note that the overlap (contributions exhibiting the same phonon wave vector) between both anti-symmetric processes depends only marginally on $\gamma$, as these processes are nearly degenerate. This means, there is always a constructive interference between P$^{43}_{12}$ and P$^{34}_{21}$ processes in the real and in the imaginary part of the scattering amplitude. Due to the mathematical structure of the Lorentzian oscillator equation, the inflection points of the real parts of the scattering amplitudes are equal to extrema of the imaginary part (indicated with blue, dashed lines). Further, there is always a sign change at the inflection points and therefore, the high wave vector contributions of the real and imaginary parts of the anti-symmetric processes will interfere constructively whereas the low- wave vector contributions will interfere destructively. This explains the right shoulders for the anti-symmetric process in all calculated spectra (Fig.~\ref{fig2}, first row). The symmetric processes, due to the large difference of phonon momenta involved, only interfere marginally with each other.\newline
As can be seen in Figs.~\ref{fig2} (a),(b) and Fig.~\ref{fig1} (a), an increase of the broadening factor $\gamma$ leads to an increase in relative intensity of the anti-symmetric scattering processes. This effect is mainly caused by the distinct resonances of the individual double-resonant scattering processes. Processes with an overall stronger resonance 
are diminished to a larger extent compared to processes with an overall weaker resonance when the broadening parameter $\gamma$ is increased. This can be seen for all symmetric P$^{33}_{22}$ and P$^{44}_{11}$ processes in Figs.~\ref{fig2} (a) and (b). Their intensity ratio decreases when $\gamma$ is increased. The dispersion of the $\pi_{1}/\pi_{1}^{*}$ bands (involved in the P$^{33}_{22}$ process as shown in Fig.~\ref{figDR}) are smaller compared to the $\pi_{2}/\pi_{2}^{*}$ bands (involved the P$^{44}_{11}$ process) resulting in a stronger resonance of the double-resonant scattering process and thus a slightly larger intensity. Although different for small $\gamma$, their distinct intensities will converge as $\gamma$ increases.\newline In anti-symmetric scattering processes, an index change of the electron/hole bands is involved. Compared to the $\pi_{1}/\pi_{1}^{*}$ bands, $\pi_{2}/\pi_{2}^{*}$ bands exhibit different dispersions, another electron-hole asymmetry, and the absolute value of the hole band offset ($\approx$\,348\,meV) is different to the electron band offset ($\approx$\,372\,meV). This results in a weaker resonance effect in the overall scattering process. As a consequence, the intensity drop of the anti-symmetric processes is systematically lower than for the symmetric processes. This explains the relative intensity gain of anti-symmetric processes when the broadening factor $\gamma$ is increased. The same holds true when the intensities of the symmetric processes $P_{11}^{44}$ and $P_{22}^{33}$ are compared: due to different dispersions of the $\pi_1/\pi_1^*$ and $\pi_2/\pi_2^*$ bands, the resonance of the $P_{11}^{44}$ process is in general broader and less intense compared to the $P_{22}^{33}$ process.\newline 
We will now discuss the case when - as in an experiment - $\varepsilon$ and $\gamma$ are increased simultaneously. As apparent in Figs.~\ref{fig2} (c) and (d), the relative intensity of anti-symmetric processes decreases for higher excitation energies although $\gamma$ increases. This effect seems to be in contrast to our previous observation. However, we will show that this effect suits into a unified picture of the double-resonance process. The main reason for this behavior is destructive interference between symmetric and anti-symmetric processes. In contrast to $\varepsilon_{\text{L}}=1.8\,$eV, the overlap for $\varepsilon_{\text{L}}=2.8\,$eV between contributions from symmetric and anti-symmetric processes in the real and in the imaginary part is larger. This in turn results in a huge intensity drop of the anti-symmetric processes in the real part of the scattering amplitude. The same accounts for the imaginary part of the scattering amplitude but to a smaller extent.\newline
Finally, large broadening factors also lead to a frequency shift of second-order scattering processes. Large broadening factors $\gamma$ increase the overlap of different scattering processes and therefore the joint resonant phonon wave vectors. Consequently, P$^{33}_{22}$ symmetric processes are shifted towards lower phonon wave vectors and the P$^{44}_{11}$ symmetric processes are shifted towards higher phonon wave vectors. We therefore generally expect a larger frequency separation between these processes for larger broadening factors. Our calculations indicate that the frequency shift induced by a large broadening factor of $\gamma=110\,$meV [compare Fig.~\ref{fig2} (b)] amounts to $\sim2\,\text{cm}^{-1}$ both for the $P_{11}^{44}$ and $P_{22}^{33}$ processes. Thus, this effect should only be measureable for large broadening factors.\newline In summary, three effects are important in order to understand the calculations in Fig.~\ref{fig2}: (\textit{i}) The almost degenerate anti-symmetric processes will always interfere constructively leading to a general increase of their contributions to the complex 2D mode in bilayer graphene.  (\textit{ii}) P$^{44}_{11}$, P$^{43}_{12}$/P$^{34}_{21}$, and P$^{33}_{22}$ processes have, due to the different slopes of $\pi_{1,2}/\pi^{*}_{1,2}$ bands, different resonance conditions and are therefore affected differently when $\gamma$ changes. (\textit{iii}) The consideration of interference between symmetric and anti-symmetric processes is crucial to understand the 2D band line shape in bilayer graphene. It leads to asymmetric line shapes of the individual contributions, a shift of the resonant phonon wave vectors, and especially for larger excitation energies a systematic decrease of the contributions from anti-symmetric scattering processes.\\ We want to mention that the used experiments performed here are only sensitive for the square of the complex scattering amplitude. Therefore, the separation of its real and imaginary parts is as an auxiliary framework for the understanding of double-resonant Raman scattering in bilayer graphene.

Next, we want to give an experimental verification of the calculational results from above. We will only discuss the general dependence of the $2D$-mode line shape as a function of $\gamma$ and will not attempt a precise quantification of the different scattering rates. We start with an analysis of the line-shape variations as a function of the laser excitation energy of the spectra alread presented in Ref.~\cite{{PhysRevLett.113.187401}}. In Figure~\ref{fig3}, the $2D$ mode of bilayer graphene measured at different laser energies in the range from 1.96\,eV to 2.81\,eV is shown. Following Eq.~(14) in Ref.~\cite{PhysRevB.84.035433}, the value of $\gamma^{\text{e-ph}}$ changes from 63\,meV at 1.96\,eV laser excitation energy to 107\,meV at 2.81\,eV laser energy. The presented spectra are normalized and shifted in frequency to the contribution from the anti-symmetric processes. We observe a strong variation of the ratio between symmetric and anti-symmetric processes, \textit{i.e.}, the intensity of the symmetric contributions drastically increases with respect to the intensity of the anti-symmetric processes. Since the measurements were performed on freestanding bilayer graphene, which is to a good approximation free of doping, defects, and strain, tuning the laser excitation energy effectively changes the contribution from electron-phonon scattering to the electronic broadening $\gamma$. We therefore experimentally observe the case shown in Fig.~\ref{fig2} (c) and (d), where only $\gamma$ and $\varepsilon_{\text{L}}$ determine the quantum interference. The experimentally observed behavior is in very good agreement with the calculations. In the experiments, the frequency difference of the symmetric P$^{44}_{11}$ process and the anti-symmetric processes diminishes when increasing the laser excitation energy. Instead, the frequency gap to the symmetric P$^{22}_{33}$ process slightly increases. On the one hand these shifts can be explained by the lower TO (transverse optical) phonon splitting for smaller phonon wave vectors at larger excitation energies~\cite{PhysRevLett.113.187401}. On the other hand, due to the systematic larger resonant phonon wave vectors involved, the local TO phonon dispersion for the symmetric P$^{44}_{11}$ processes is larger compared to the symmetric P$^{33}_{22}$ processes\cite{PhysRevLett.113.187401}. This explains the slightly different shift rates as indicated by arrows in Fig.~\ref{fig3}.

\begin{figure}
\centering
\includegraphics[width=0.93\columnwidth]{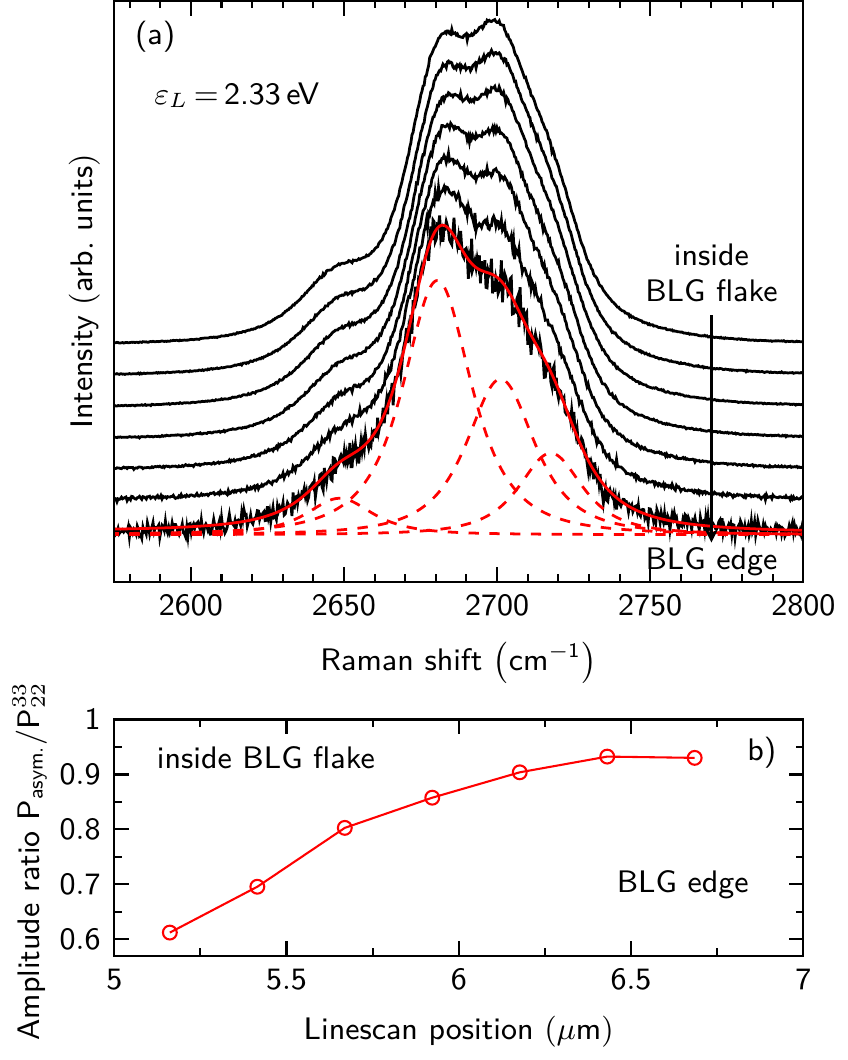}%
\caption{Raman spectra of a line scan across a bilayer graphene edge. The lowest spectrum refers to the bilayer edge, whereas the topmost spectrum was measured inside the bilayer graphene flake. The solid re line denotes a fit with four Baskonian profiles \cite{PhysRevB.78.125418}, the dashed lines represent the single contributions. Spectra are normalized to the same $P_{\text{asym.}}$ intensity and vertically offset for clarity. (b) Evolution of the ratio between P$^{33}_{22}$ and the contribution from anti-symmetric processes along the line scan across the bilayer graphene edge.
 \label{fig:BilayerDefects2}}%
\end{figure}  
As theoretically and experimentally discussed in Refs.~\cite{PhysRevB.84.035433} and \cite{PhysRevB.91.205413}, the electronic broadening does also depend on the defect concentration in the graphene sample. In general, a larger defect concentration leads to a higher electron-defect scattering rate and thus to an increased value for $\gamma^{\text{D}}$. In fact, $\gamma^{\text{D}}$ should be directly proportional to $n_D$ \cite{PhysRevB.84.035433}. However, $\gamma^{\text{D}}$ is usually significantly smaller than $\gamma^{\text{e-ph}}$, \textit{e.g.}, $\gamma^{\text{e-ph}} = 63$\,meV for $\varepsilon_{\text{L}}$=1.96\,meV and $\gamma^{\text{D}} = 7$\,meV in defective single-layer graphene at a defect concentration of $n_D = 0.9\times10^{12}$\,cm$^{-2}$ \cite{PhysRevB.91.205413,PhysRevB.84.035433}. In the present study, we use ion-irradiated graphene samples that were exfoliated on standard silicon substrates with 100\,nm SiO$_2$. Ion-irradiated graphene samples allow a precise quantification of the average defect distance from the ion flux. This approach is independent from geometrical parameters of the defect and more reliable than a determination from the measured $D/G$-mode ratio. In Figure~\ref{fig:BilayerDefects}\,(a), we present Raman spectra of the $2D$ mode in defective bilayer graphene with three different average defect distances $L_D$. We observe a dependence of the ratio between symmetric and anti-symmetric contributions on the defect concentration, \textit{i.e.}, with increasing defect concentration (decreasing average defect distance) we observe an increase of the ratio between anti-symmetric and symmetric processes [compare Fig.~\ref{fig:BilayerDefects}\,(b)]. Again, this is in good agreement with our theoretical predictions from Fig.~\ref{fig1}\,(a). An increasing defect concentration leads to a larger value for the electronic broadening, which in turn results in an increase of the contribution from anti-symmetric scattering processes compared to symmetric processes. The small relative intensity variations can be explained by the small contribution of $\gamma_{D}$ to the total broadening $\gamma$. Our findings should also be applicable for freestanding graphene since the total value of $\gamma_{D}$ does not depend on the substrate.\\ As a last point we want to show Raman spectra where the line shape variations of the 2D mode in bilayer graphene can be observed in a simple experiment. Raman spectra in Fig.~\ref{fig:BilayerDefects2} were recorded while performing a line scan across the edge of a bilayer flake. As can be seen, we observe  a strong variation of the 2D mode line shape as a function of the position, \textit{i.e.} at the edge we observe an increased ratio between anti-symmetric and symmetric scattering processes as compared to the inside of the bilayer flake. The edge can be seen as a distortion of the translational symmetry of the bilayer graphene lattice and therefore it can been seen as a defect comparable to those in Fig.~\ref{fig:BilayerDefects}. In addition, graphene edges are most likely compressively strained\cite{PhysRevLett.102.166404,PhysRevLett.101.245501} further increasing the electronic broadening. As a consequence, the broadening $\gamma$ changes as a function of the position and explains the differences in the Raman spectra.\newline The measurements as shown in Fig.~\ref{fig:BilayerDefects2} and Fig.~\ref{fig:BilayerDefects} agree well with our findings in the calculations and show how the broadening factor $\gamma$ can be tuned experimentally. \newline Another possible route to tune the discussed interference is ultra-violet Raman spectroscopy. As shown in Ref.~\cite{PhysRevB.92.041401}, laser excitation energies near the $M$ point transition in graphene, \textit{i.e.} around 4.7\,eV and above, suppress the most dominant, so-called "inner" scattering paths for the 2D mode in graphene or graphite. The different $\pi_{1}\rightarrow\pi_{1}^{*}$ and $\pi_{2}\rightarrow\pi_{2}^{*}$ $M$ point transition energies in bilayer graphene might be used to address single scattering processes individually by carefully choosing the laser excitation energy. With this approach, it should be possible to turn off the interference between symmetric and anti-symmetric processes. This might give further insights into the quantum interference in bilayer graphene.

In conclusion, we showed the influence of quantum interference to the complex 2D Raman mode in bilayer graphene. We analyzed both theoretically and experimentally how the broadening factor $\gamma$ tunes the quantum interference of different scattering processes. We find that (\textit{i}) independent from $\gamma$, there is always constructive interference of anti-symmetric processes in bilayer graphene. (\textit{ii}) Constructive and destructive interference of the real and imaginary part from the scattering amplitudes lead to asymmetric line shapes of distinct contributions of the complex 2D line shape. (\textit{iii}) Different scattering paths in bilayer graphene exhibit difference resonances. Hence, tuning the broadening factor $\gamma$ affects the scattering amplitudes differently. Intensities from the symmetric contributions to the 2D mode are generally damped to a greater extent than those from anti-symmetric contributions when $\gamma$ is increased. (\textit{iv}) For large excitation energies and therefore large $\gamma^{\text{e-ph}}$, a strong destructive interference between symmetric and anti-symmetric contributions occurs. This leads to a strong damping of the symmetric scattering processes in bilayer graphene. It even exceeds the effect pointed out in (\textit{iii}). (\textit{v}) Due to slightly changed resonances, interference effects lead to a frequency downshift of the P$^{44}_{11}$ contributions and an upshift of the P$^{33}_{22}$ contributions. Furthermore, we presented how the electronic broadening factor $\gamma$ can be experimentally tuned by defects and edges. 
   
\vspace{0.5cm}

\section*{Acknowledgments}
F.H., C.T., and J.M. acknowledge financial support from the DFG under grant number MA 4079/6-2 within the Forschergruppe 1282, under grant number MA 4079/7-2 within the Schwerpunktprogramm SPP 1459, and from the European Research Council (ERC) Grant No. 259286. O.O. and M.S. acknowledge financial support from the DFG within the SPP 1459 "Graphene" under SCHL 384/15-1. Sample irradiation was performed at the IRRSUD beamline of the GANIL (Caen, France). We thank H. Lebius for his kind support. The authors further thank Matteo Calandra, Michele Lazzeri, and Francesco Mauri for providing the \textit{GW}-corrected electronic bandstructure and TO phonon dispersions.

\bibliographystyle{apsrev4-1}

\end{document}